# Spin-Torque-Induced Rotational Dynamics of a Magnetic Vortex Dipole

G. Finocchio[1], O. Ozatay[2,3], L. Torres[4], R. A. Buhrman[3], D.C. Ralph[3], B. Azzerboni[1]

[1]Dipartimento di Fisica della Materia e Tecnologie Fisiche Avanzate. University of Messina. Salita Sperone 31, 98166, Messina Italy.
[2]Hitachi GST, San Jose Res. Ctr., San Jose, CA, 95135
[3]Cornell University, Ithaca, New York 14853-2501, USA.
[4]Departamento de Fisica Aplicada. Universidad de Salamanca. Plaza de la Merced s/n 37008. Salamanca. Spain.

**We study, both experimentally and by numerical modeling, the magnetic dynamics that can be excited in a magnetic thin-film nanopillar device using the spin torque from a spatially localized current injected via a 10s-of-nm-diameter aperture. The current-driven magnetic dynamics can produce large amplitude microwave emission at zero magnetic field, with a frequency well below that of the uniform ferromagnetic resonance mode. Micromagnetic simulations indicate that the physical origin of this efficient microwave nano-oscillator is the nucleation and subsequent steady-state rotational dynamics of a magnetic vortex dipole driven by the localized spin torque. The results show that this implementation of a spintronic nano-oscillator is a promising candidate for microwave technology applications.**

# I. INTRODUCTION

Spin-transfer torque,[1,2] by which spin-polarized currents can be used to reorient magnetic devices as an alternative to using magnetic fields, offers the possibility of applications that include magnetoresistive random access memory, nano-oscillators, and radiofrequency detectors. The mechanism of the torque and the resulting magnetic dynamics have been studied in several different sample geometries, including multilayer point-contacts,[3,4,5,6,7] spin valves,[8,9,10,11,12,13,14] magnetic tunnel junctions,[15,16,17,18] and magnetic wires in which spin torque moves domain walls.[19,20,21,22,23] To control device performance, it is critical to understand the nature of the spin-torque-driven magnetic excitations. Previous measurements in nanopillar spin valve systems with spatially-uniform direct current injection showed that large-amplitude coherent magnetic dynamics can be excited over a wide range of bias currents with a fundamental frequency near the natural ferromagnetic resonance frequency of the (approximately) uniform mode (several GHz).[11,24] Spin torque has also been shown to be able to excite steady-state magnetization oscillations in a strongly non-uniform magnetic configuration, namely a vortex state in a spin-valve nanopillar device.[25] The magnetic orientation of the vortex core can be reversed by spin torque from an oscillating current,[26] and predictions suggest that a direct current may also be able to achieve this result.[27]

Here, we report by means of frequency-domain measurements on spin valve devices near zero magnetic field that a strong localized spin torque caused by a spatially non-uniform injection of current in a spin valve device can induce microwave signals with frequency well below the uniform ferromagnetic resonance mode. Based on micromagnetic simulations, we suggest that the current is able to nucleate and subsequently drive the steady state rotation of a vortex dipole around the current-injection site.[28,29] (A vortex dipole is a vortex-antivortex (VA)

pair.[29]) Our geometry differs from previous experiments on spin-torque-driven steady-state magnetic oscillations in 10s-of-nm-diameter point contact devices[6,30,31] in that previous work explored excitations within a ferromagnetic layer whose lateral extent was effectively infinite, while we study excitations within a finite 150 nm × 250 nm layer in response to current applied through a 10s-of-nm-diameter aperture. Our experimental observations are, however, similar in some respects to the results on the low-frequency modes in References 31 and 32, which were ascribed to the perturbation of magnetic vortex states.[31,32,33] Other micromagnetic simulations of spin-torque-driven magnetic dynamics in point contact devices have considered only much higher frequency modes present at larger magnetic fields.[34,35]

## II. SAMPLE FABRICATION AND MEASUREMENT PROCEDURES

Our devices are fabricated[36] using thin film deposition and multiple layers of aligned electron-beam lithography to enable the local injection of spin-polarized currents within a nanopillar spin valve. The device layer structure is Py($Ni_{81}Fe_{29}$) (5 nm) / $Al_2O_3$ (3.5 nm) / Cu (8 nm) / Py (20 nm) (see Fig. 1(a)). Before deposition of the Cu layer, we etch a 20-40 nm diameter aperture halfway through the $Al_2O_3$ layer using ion milling. The sample is then transferred to a sputter deposition chamber equipped with an argon ion mill (base pressure $5 \times 10^{-8}$ Torr), which is used to complete the excavation of the aperture down to the surface of the top of the thinner Py layer, to provide the current path for local injection. Without breaking vacuum, the rest of the layer structure is then deposited. After completion of the layer deposition, the device is ion-milled into a nanopillar geometry with an approximately elliptical cross sectional area (250 nm × 150 nm major and minor diameters), with the nano-aperture located approximately at the center of the ellipse. Both magnetic layers are etched through in defining the outer sidewalls of the nanopillar. We will refer to the thinner Py layer as the free layer (FL) and to the thicker layer as the pinned layer (PL). We will employ a Cartesian system of coordinates in which the major axis of the ellipse is the x-axis ($\hat{x}$) and the in-plane hard axis is the y-axis ($\hat{y}$), and we will use the convention that positive current corresponds to the flow of electrons from the FL to the PL. We have previously reported the spin-torque switching properties of devices from the same fabrication batch.[36]

For frequency-domain studies, we apply a dc current bias flowing perpendicular to the sample layers. Because the resistance is dominated by the region in the vicinity of the nanoaperture, the measurements are sensitive only to magnetoresistance signals from magnetization oscillations near this contact region. At zero field and current, the device configuration is an antiparallel alignment of the magnetizations in the FL and PL, due to their magnetostatic coupling. The measurements are performed at a background temperature of 5.6 K and for static magnetic fields applied in-plane along the easy axis of the sample (Fig.1(a)). We observe oscillatory microwave signals due to time-dependent resistance oscillations in response to a dc current in the range -2 to -5 mA (we did not explore beyond -5 mA) when the electrons flow from the PL to the FL (only for negative current) and for a narrow range of fields near zero field (-25 mT to 20 mT). In other parts of the dynamical phase diagram, at larger magnitudes of current and magnetic field, we also observe other higher-frequency dynamical states qualitatively similar to previous observations.[6,11] In this paper we will discuss only the more-novel lower-frequency oscillations present near zero field.

### III. MEASUREMENTS

Figure 1(b) shows the frequency spectrum of the microwave signal produced by the sample at -4 mA dc current bias and zero applied field, as measured with a spectrum analyzer.[11,36] The signal is characterized by a single mode, with the largest spectral peak (P1) at the fundamental frequency and smaller peaks (P2, P3, etc.) at higher harmonics. The fundamental peak lies in the range 0.7-0.9 GHz, much less than the ferromagnetic resonance frequency corresponding to approximately spatially uniform precession (4-6 GHz). As can be seen in Fig. 1(c), once the applied current magnitude is large enough to drive the persistent oscillations, the frequencies of the fundamental mode and harmonics are essentially insensitive to further increases in current magnitude. Figure 1(d) and (e) show the frequency (circles 'o') and linewidth (full width at half maximum (FWHM)) (plus '+') of the fundamental peak as a function of current bias $I$ (at $H = 0$ mT) and applied field $H$ (at $I = -4$ mA) respectively. After the turn on of persistent oscillations at about -2.5 mA, the frequency varies by less than 4% to -5 mA. As a function of increasing $H$ (at $I = -4$ mA) the frequency shifts down by about 9% between -25 mT and 20 mT, and for positive magnetic fields (approaching the AP-to-P transition) the dependence on $H$ is particularly weak. As noted above, no magnetization dynamics are measured for applied fields smaller than -25 mT or larger than 20 mT; the micromagnetic simulations indicate that these threshold fields are close to transitions to the AP and P states respectively. The most striking feature of the microwave spectra is that the peaks are quite sharp. The linewidth of the fundamental peak decreases as a function of increasing current and stabilizes at 4 MHz for currents more negative than -3.5 mA (Fig. 1(d)). Away from the switching field thresholds there is a region of constant linewidth in the field range -2.5 to 12.5 mT (Fig. 1(e)). At both higher and lower fields, the linewidth increases by up to a factor of 20.

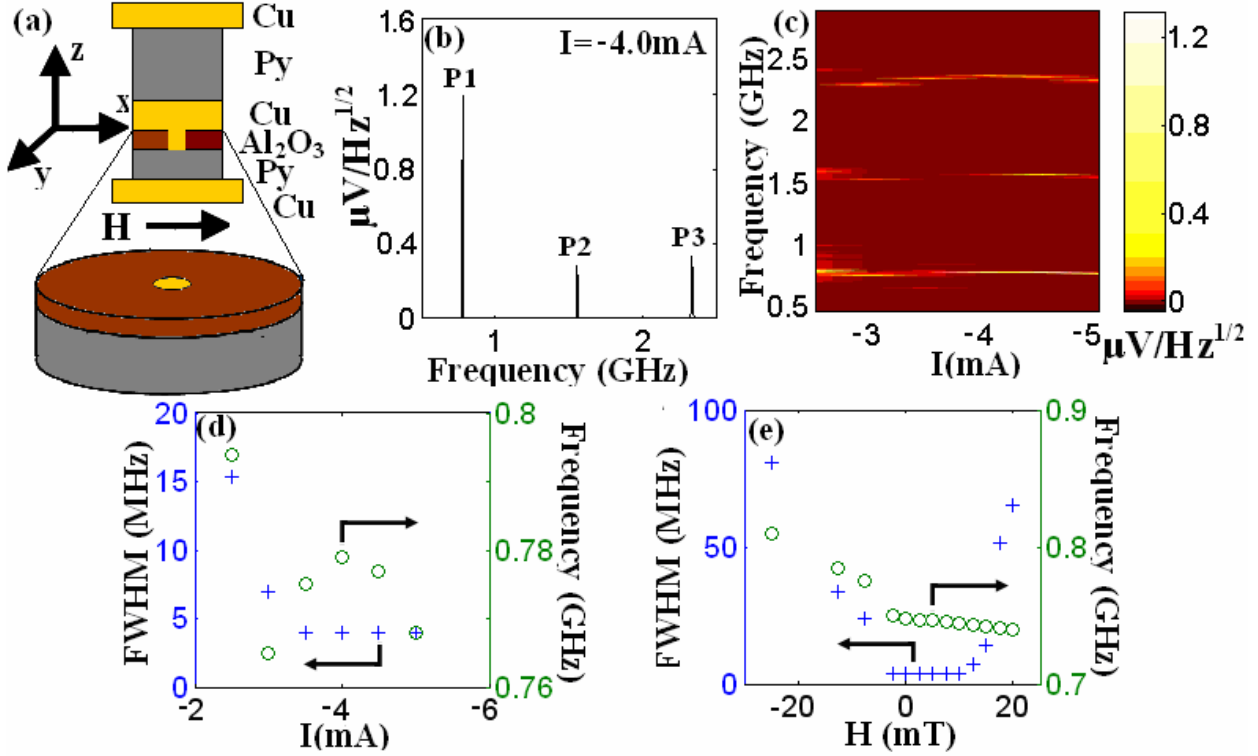

FIG. 1. (Color online) Measurements of dc-driven magnetization dynamics for sample 1 at a background temperature of 5.6 K: (a) schematic of the device geometry. (b) microwave voltage spectrum driven by a current of -4 mA (no applied magnetic field) where fundamental mode (P1) of the free layer and harmonics (P2, P3) can be observed. (c), experimental voltage spectra as function of current (no applied magnetic field). (d) and (e), experimental frequency (o) and FWHM linewidth (+) as function of current (no applied magnetic field) (d), and as function of field at $I = -4$ mA (e).

## IV. MICROMAGNETIC SIMULATIONS AND DISCUSSION

To understand the nature of these magnetization dynamics driven by spatially non-uniform current injection, we performed a systematic numerical analysis by means of micromagnetic simulations of the Landau-Lifshitz-Gilbert (LLG) equation with the Slonczewski torque term for the device structure studied experimentally.[37] The simulations have been performed for various nano-aperture diameters $d$ (30 - 60 nm), and aspect ratios of the elliptical nanomagnet at both 0 K and at temperatures selected to model local heating. The details of the simulation procedures and parameters are given in the Appendix.

The simulations suggest that the experimentally observed magnetization dynamics are due to the presence of a vortex-antivortex (VA) pair which at steady state rotates persistently around the current injection site, giving rise to resistance oscillations. Figure 2 shows representative snapshots of magnetization configurations in the FL from micromagnetic simulations at steady state in which the VA-pair rotates counterclockwise (see supplementary video 1[38]). The arrows indicate the direction of the in-plane magnetization and the color code is for the out of plane ($m_z$) component of the magnetization (blue negative, red positive). It can be seen that the core regions of both vortices and antivortices contain a significant out-of-plane component. We describe this be defining the polarity of the vortex ($p_V$) and the antivortex ($p_A$) to be +1 for positive $m_z$ and -1 for negative $m_z$. The microwave signal is due to the oscillation of the magnetization in the FL below the nano-aperture which moves on account of the rotation of VA pair. With no thermal effects, we find that the sense of rotation of the VA pair can be either clockwise or counterclockwise (both of these configurations are stable). The sense of rotational motion is related to the polarities of the vortex and the antivortex, this motion is clockwise for $p_A = +1$ and $p_V = -1$, counterclockwise for $p_A = -1$ and $p_V = +1$.[39] During the course of the rotation, the chirality with which the instantaneous magnetization configuration winds around the core region of the vortex can evolve from clockwise (Fig. 2 snapshot 2) to counterclockwise (Fig. 2 snapshot 4) and vice versa, with dynamics similar to those described in Ref. 40. If we take local heating into account in the micromagnetic simulations, we find that the system can also switch polarities and thereby jump between clockwise and counterclockwise rotational states.

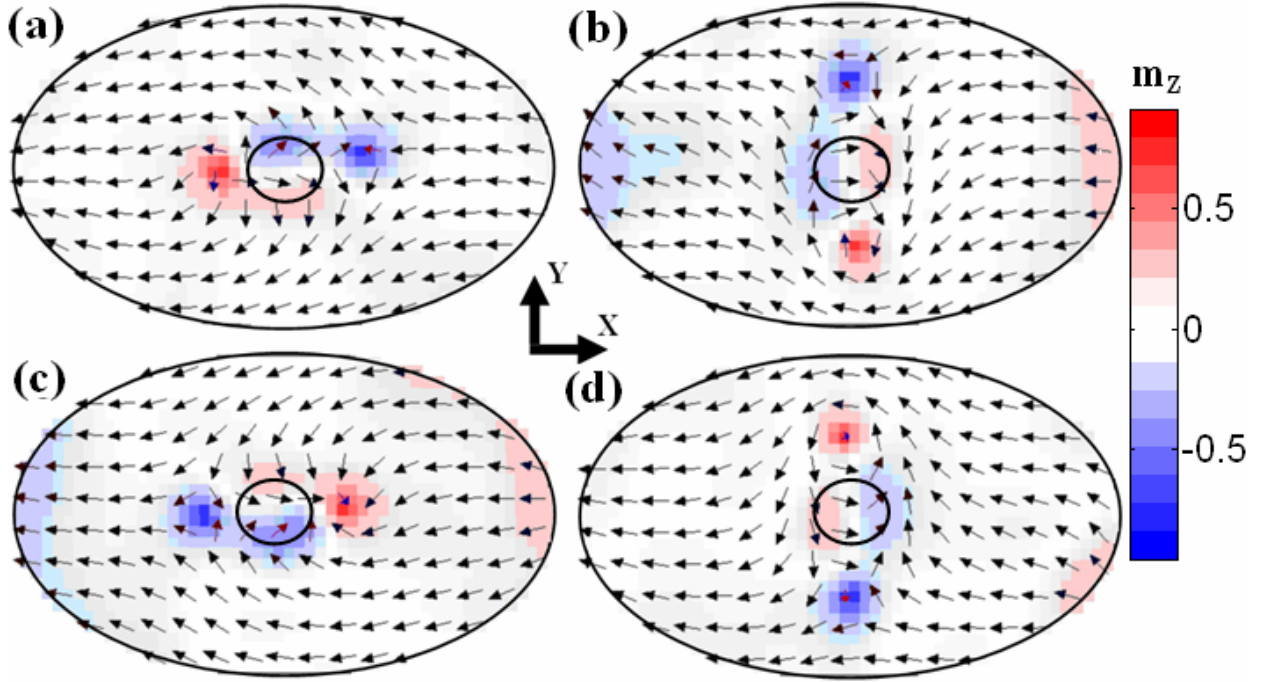

FIG. 2. (Color online) Snapshots of the simulated magnetization dynamics for a VA pair rotating under the influence of a dc current at zero temperature. The simulation is for a nano-aperture of diameter $d = 40$ nm (the circle represents the injection site) for a current $I = -4.5$ mA at $H = 0$ mT. The arrows indicate the direction of the in-plane magnetization, the color indicates the amplitude of the z component. The sense of rotational motion (counterclockwise) is identified by the sequence of the letters.

To evaluate the frequency of the oscillations in the simulations, we calculated the spectrum of the average y-component of the magnetization in the free layer. (This quantity is not meant to model the resistance oscillations, but only to characterize the frequency of motion[41]). The main panel of the Fig. 3(a) displays the normalized spectrum computed for $I = -4$ mA, $d = 50$ nm and $H = 0$ mT. The numerical calculations reproduce the main qualitative features of the experimental observations, including the weak dependence of the oscillation frequency on bias current (Fig. 3(a) upper inset and Fig. 3(b)), the decrease in frequency with increasing magnetic field (Fig. 3(a) lower inset and Fig. 3(c)), and the presence of oscillations only for negative currents. As in the experiments, the oscillations in the simulations are present only in a narrow range of magnetic field about $H = 0$ mT (-25 to 15 mT for $d = 40$ nm), outside of which the VA pair is annihilated. Quantitatively, the precession frequency in the simulation is larger than in the experiment by 25-30% and the dependence of the frequency on magnetic field is somewhat stronger in the simulation than in the experiment.

In the simulations of the vortex-antivortex rotation, the magnetic moment below the nano-aperture can precess to angles as large as 80% of full reversal, consistent with the large-amplitude resistance oscillations observed in the measurements. The experimentally-measured threshold current to excite coherent microwave oscillations, -2.5 mA (at zero applied field), corresponds to an aperture size in the range of 40-50 nm in the micromagnetic simulations. For a fixed current, increasing the contact diameter causes the oscillation frequency to decrease monotonically because the distance between vortex and antivortex increases and the fundamental frequency is governed by this effective distance. In Fig. 3(a), we also show the spatial distribution of the contributions to each of the spectral peaks computed by means of the micromagnetic spectral mapping technique (MSMT)[42,43] for an applied current of $I = -4.0$ mA and a nanocontact diameter $d = 50$ nm (power increases from white to black). The power in the fundamental peak originates from a region centered about the perimeter of the aperture, as might be expected for a rotating VA pair.

We have previously reported micromagnetic simulations for a different point-contact spin-valve geometry which found gyrotropic oscillations of a single magnetic vortex, rather than vortex-antivortex rotation.[33] We suggest that these two scenarios can be distinguished experimentally in that single-vortex oscillations have a more-strongly current-dependent frequency. We find that a vortex state can be more energetically favourable for very large Oersted fields (larger currents), for an open geometry in which the magnetic layers are effectively infinite in extent (like previous point contact devices[29,30]), or for thicker layers (e.g., the 60 nm layer in ref. [23]). In the case we consider here of a thin free layer (5 nm) patterned within a nanopillar geometry, the vortex dipole state provides a low-energy compromise between the Oersted field and the magnetostatic energy, because the inner region of the vortex dipole provides a good accommodation to the Oersted field while the approximately spatially uniform outer region accommodates the coupling field from the pinned layer (35 mT along the $-\hat{x}$ direction). In addition, because the vortex and anti-vortex have cores that point in opposite directions (see Fig. 2), the vortex dipole gives a lower magnetostatic energy than a simple vortex state for our thin magnetic free layer.

The frequency of a rotating VA pair has been computed analytically for a thin-film sample that is infinite in lateral extent:[28,29]

$$f = \frac{4\gamma_0 A}{\mu_0 M_S d_{VA}^2}, \qquad (1)$$

where $d_{VA}$ is the distance between the vortex and the antivortex, $A$ is the exchange constant, $\gamma_0$ is the gyromagnetic ratio, and $M_S$ is the saturation magnetization. We would not expect Eq. (1) to be accurate for our samples, since we are considering VA rotations in devices that are confined in the lateral direction and which have significant demagnetization fields that break cylindrical symmetry. Nevertheless, when we use appropriate values of $A = 1.3 \times 10^{-11}$ J/m and $M_S = 650 \times 10^3$ A/m for our Py free layer, we find that if we scale the right hand side of Eq. (1) by the factor 0.5 the expression gives a good description of the frequency found by our micromagnetic simulations, with $d_{VA}$ computed directly from the spatial configuration of the magnetization. For example, for $I = -4.0$ mA, $H = 0$ mT, and $d = 50$ nm (Fig. 3(a)) the simulation gives $d_{VA} = 80$ nm and a fundamental precession frequency of 1.12 GHz, compared to $f = 2.20$ GHz for Eq. (1).

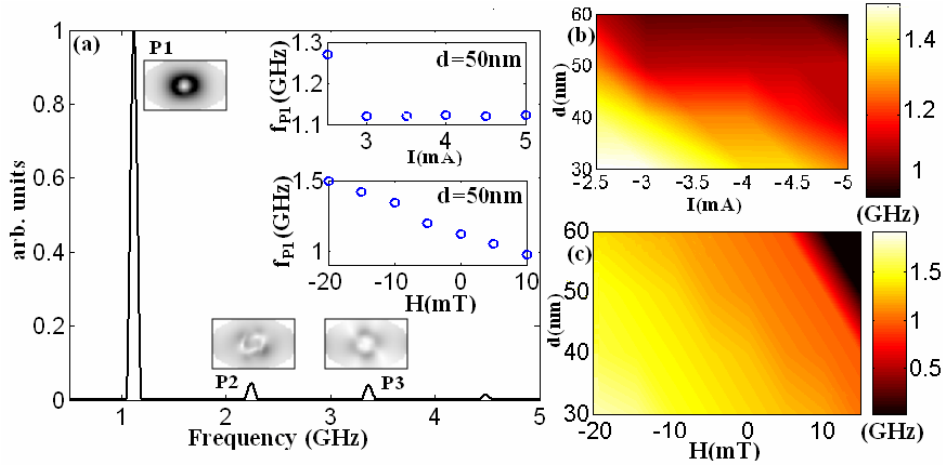

FIG. 3. (Color online) Results of the micromagnetic simulations. (a) Frequency spectrum for $I = -4$ mA, $d = 50$ nm (with $H = 0$). The picture insets show the spatial distribution of the sample regions contributing to the power (increasing from white to black) in each of the spectral peaks as computed by means of the MSMT technique referenced in the text. Upper inset: computed fundamental frequency versus current (for $H = 0$ mT, $d = 50$ nm). Lower inset: computed frequency versus applied magnetic field (for $I = -4$ mA, $d = 50$ nm). (b) Computed fundamental oscillation frequency as function of the current and the diameter of the nano-aperture (for $H = 0$ mT). (c) Computed fundamental oscillation frequency as function of magnetic field and the diameter of the nano-aperture (for $I = -4$ mA).

If we take into account the thermal effects in the micromagnetic simulations by assuming that Ohmic heating raises the sample temperature to 90 K (for $I = -2.5$ mA, see the Appendix), the magnetization dynamics becomes noisy (Fig. 4(a)), but the VA-pair dynamics are preserved. In particular we obtain the following theoretical results: (i) the oscillation frequency is essentially independent of the local temperature; (ii) the thermal field tends to annihilate the VA pair for large contact diameters (for the device parameters used in this calculation the thermal annihilation occurs for $d \geq 50$ nm); (iii) transitions are often seen between dynamical states in which the VA-pair rotates in clockwise or counterclockwise sense. Figure 4(b) shows examples of this type of mode transition at the times $t_1$ and $t_2$; for times smaller than $t_1$ or larger than $t_2$ the sense of rotation is clockwise, whereas for times between $t_1$ and $t_2$ the sense of rotation is counterclockwise. (see supplementary video $2^{38}$ for the jump at $t_1$). Thermally-induced mode transitions and thermal deflection about the equilibrium magnetic trajectory (thermal fluctuations) can both be expected to broaden the linewidths of the emitted microwaves as a function of increasing temperature.

We have found that the simulations of vortex-antivortex rotation provide an excellent qualitative explanation of the measured microwave oscillations in the nano-aperture nanopillar devices, but there are quantitative differences in the oscillation frequency at the 25-30% level. We have considered the potential influence of several factors which might alter the frequency at this level, including (i) misalignment in the nano-aperture location with respect to the center of the ellipse, (ii) the presence of ion milling damage in the vicinity of the nano-aperture, (iii) spatial variations in the current density near the nano-aperture, and (iv) variations in the device aspect ratio from the idealized shape assumed in the simulation. (i) The alignment accuracy of our electron-beam lithography is approximately 10 nm. The micromagnetic simulations show that a shift of the nano-aperture from the center of the elliptical cross section by 10 nm (in both the x and y directions) would decrease the oscillations frequency by approximately 5-10 % relative to the ideal case. (ii) We performed a systematic micromagnetic study of the effect of the ion-mill damage (modeled as isolated cells below the nano-contact that are metallic but non-magnetic) as a function of location and size. We observe that the VA rotational dynamics are present when the defect area is smaller than 4 computational cells ($\leq 75$ nm$^2$). The frequency spectra are sensitive to the location of the defects due to a change in the VA pair motion (see supplementary video 3[38]). (iii) Recent magnetic force microscopy studies of electron flow mapping showed that electron transport around defects in conductors is highly dependent on the local geometry of the defect, with the potential for significant fluctuations in current density near defects.[44] To qualitatively study the effect of variations in the current distribution we computed the VA pair rotation frequency using both a perfectly uniform current density distribution below the nano-aperture and the distribution computed by solving the Poisson equation (see the Appendix). The frequencies differed by 30%, with the spatially-uniform case giving the larger frequency. (iv) The micromagnetic simulations show that changes in the aspect ratio of the ellipse can also cause changes in the oscillation frequency of the excited mode. For a smaller aspect ratio (less circular devices) the frequencies of the excited modes are somewhat larger. We estimate that uncertainties in the device dimensions introduce a maximum uncertainty of 20% in the values of the oscillation frequency computed in the simulations. Another factor that is not included in the simulations but could be important experimentally is magnetization dynamics in the PL[25] that couple to the FL dynamics.

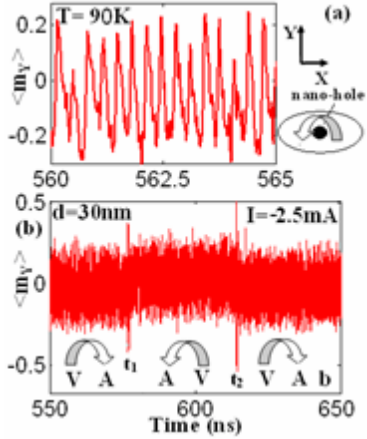

FIG. 4. (Color online) Temporal evolution of the y-component of the average magnetization for $I = -2.5$ mA, $d = 30$ nm at $T = 90$ K. (a) Close-up for a short time window. (b) A long time windows within which there are jumps between clockwise and counterclockwise VA pair rotation at times $t_1$ and $t_2$. The arrows show the sense of the rotation of the VA pair.

## V. CONCLUSIONS

Our measurements show, for applied magnetic fields near zero, that spatially non-uniform injection of a dc spin-polarized current onto a nanoscale ferromagnet can produce strong resistance oscillations with frequency much lower than the ordinary ferromagnetic resonance frequency for uniform precession, with narrow linewidths, and with only weak dependence of the oscillation frequency on the current and field values. Micromagnetic calculations demonstrate that these observations can be explained by spin-torque-driven steady-state rotation of a vortex-antivortex pair. This result opens new strategies for the design of spintronic nano-oscillators, in that the VA rotation mode allows magnetization dynamics to be excited at zero applied magnetic field, and the frequency of the VA rotation mode can be tuned by varying the aperture diameter or the sample aspect ratio while maintaining a narrow linewidth.

ACKNOWLEDGMENTS

The authors thank Antonino Barresi and Antonino Romeo for software support and Vlad Pribiag, Stravos Komineas, and Giancarlo Consolo for very helpful discussions and comments. This work was supported in part the Office of Naval Research, the NSF/NSEC program through the Cornell Center for Nanoscale Systems and an IBM-Faculty Partnership award. The work was carried out in part at the Cornell NanoScale Facility, a member of the National Nanotechnology Infrastructure Network, which is supported by the National Science Foundation (Grant ECS 03-35765), and it benefited from use of the facilities of the Cornell Center for Materials Research, which is supported by the NSF/MRSEC program. This work was partially supported by projects MAT2005-04827 from Spanish government.

APPENDIX: DESCRIPTION OF THE MICROMAGNETIC SIMULATIONS

Our micromagnetic simulations are based on the solution of the Landau-Lifshitz-Gilbert-Slonczewski (LLGS) equation:[1,45,46,47]

$$\frac{d\boldsymbol{m}}{d\tau} = -\boldsymbol{m} \times \boldsymbol{h}_{\text{eff}} + \alpha \boldsymbol{m} \times \frac{d\boldsymbol{m}}{d\tau} - \frac{g|\mu_B||j_Z(x,y)|}{e\gamma_0 M_S^2 L}\varepsilon(\theta)\boldsymbol{m} \times (\boldsymbol{m} \times \boldsymbol{m}_p) \qquad (A1)$$

where $\alpha$ is the damping coefficient, $g$ is the gyromagnetic splitting factor, $\gamma_0$ is the gyromagnetic ratio, $\mu_B$ is the Bohr magneton, $\alpha$ is the Gilbert damping, $j_Z(x,y)$ is the current density computed as described below, $L$ is the thickness of the free layer, $e$ is the electron charge, $\boldsymbol{m} = \boldsymbol{M}/M_S$ is the normalized magnetization of the free layer, $\boldsymbol{m}_p = \boldsymbol{M}_p/M_S$ is the normalized magnetization of the pinned layer, $M_S$ is the saturation magnetization, $d\tau = \gamma_0 M_S dt$ is the dimensionless time step, $\varepsilon(\theta)$ characterizes the angular dependence of the Slonczewski spin torque term, and $\boldsymbol{h}_{\text{eff}}$ is the effective field.[47] The material parameters we use are: an exchange constant $A = 1.3 \times 10^{-11}$ J/m, a Gilbert damping coefficient $\alpha = 0.01$, and saturation magnetization $M_S = 650 \times 10^3$ A/m. We employ a spin-torque efficiency function of the form $\varepsilon(\theta) = k/(1+b\cos(\theta))$, where $\theta$ is the angle between the magnetizations of the PL and the FL ($\cos(\theta) = \boldsymbol{m} \cdot \boldsymbol{m}_p$), and $k = 0.54$ and $b = 0.6$ are determined by fitting to experimental switching currents at positive and negative bias, as described in Ref. 36. We take into account the spatial dependence of $\varepsilon(\theta)$ by computing this value for each computational cell, although we do not attempt to incorporate lateral spin diffusion within the sample layers. The simulation does account for the magnetostatic coupling between the PL and the FL and the Oersted field due to the current.

We model the spatial distribution of the current density numerically solving the Poisson equation for a model geometry in oblate spherical coordinates.[48] For a current flowing throw a circular aperture of radius $a$ under the influence of a voltage $V$, the contours of constant electric potential are given by $\phi(\boldsymbol{r}) = \pm 0.5V\left[1-(2/\pi)\tan^{-1}(1/\xi(\boldsymbol{r}))\right]$, where $\boldsymbol{r} = x\hat{x} + y\hat{y}$, $r^2 = x^2 + y^2$, and $\xi$ can be computed by the equation $r^2/a^2 = (1+\xi^2)\left[1-z^2/(\xi^2 a^2)\right]$, where the two signs refer to the two sides of the contact. Assuming Ohm's law $j_z = \sigma E_z = (\sigma/e)\delta\phi/\delta z$ ($\sigma$ and $e$ are the conductance and the electron charge), we compute the current density distribution $j_z$ as a function of lateral position at a fixed value of $z$. Our simulations have been performed considering the current density computed at $z = -2.5$ nm. The presence of nano-apertures with different diameters is taken into account by varying the radius of the current density distribution.

We include thermal effects by adding a random thermal field to the deterministic effective field.[49,50] We assume that the fluctuating field is independent of the spin torque, the spin torque does not depend on the thermal field, and the magnetization configuration of the PL does not depend on temperature.[51,52,53] Due to Joule heating, the temperature of the sample can be higher than the background substrate temperature. We assume that the local temperature at the construction site is given by $T_S^2 = T^2 + C_V V^2$ where $T$ and $T_S$ are background and sample temperatures, respectively, $V$ is the voltage, and $C_V$ = 3.2 K/mV is a constant determined by fits to thermally-activated switching rates in the same devices. For converting between voltage and current, the typical experimental lead resistance is 20-30 Ω. To study the effects of temperature we performed simulations of 1000 ns.

We use a time step of 40 fs in our calculations. Tests performed with a time step of 30 fs gave the same results. We consider for the discretization cubic cells of 5 nm on a side. Simulations performed with smaller cells of (2.5 × 2.5 × 5.0 nm$^3$) exhibit changes in the oscillation frequency of less than 4 %.